\newcommand{\ZZ}{\mathbb{Z}_2}
\newcommand{\CC}{\mathbb{C}}
\newcommand{\Gr}{Gr\"obner }

\documentclass[a4paper]{spie}  

\usepackage{graphicx}
\usepackage{amsmath}
\usepackage{amssymb}

\title{An Algorithm for Constructing Polynomial Systems Whose Solution Space
Characterizes Quantum Circuits}

\author{Vladimir P. Gerdt and Vasily M. Severyanov
\skiplinehalf Laboratory of Information Technologies \\
Joint Institute for Nuclear Research \\ 141980 Dubna, Russia }

\authorinfo{Further author information: (Send correspondence to Vasily M. Severyanov)\\
Vladimir P.Gerdt: E-mail: gerdt@jinr.ru, Telephone: +7 (09621) 63437\\
Vasily M. Severyanov: E-mail: severyan@jinr.ru, Telephone: +7 (09621) 64800}

\begin{document}
\maketitle

\begin{abstract}
An algorithm and its first implementation in C\# are presented for
assembling arbitrary quantum circuits on the base of Hadamard and
Toffoli gates and for constructing multivariate polynomial systems
over the finite field $Z_2$ arising when applying the Feynman's
sum-over-paths approach to quantum circuits. The matrix elements
determined by a circuit can be computed by counting the number of
common roots in $Z_2$ for the polynomial system associated with the
circuit. To determine the number of solutions in $Z_2$ for the
output polynomial system, one can use the \Gr bases method and
the relevant algorithms for computing \Gr bases.

\end{abstract}


\section{INTRODUCTION}
\label{sect:introduction}

When applying the famous Feynman's sum-over-paths approach to a quantum circuit, as it was shown
in~\cite{Dawson}, one can associate with an arbitrary quantum circuit a polynomial system over the
finite field $\ZZ$.  For $N$ input qubits the system contains $N+1$ polynomials: each
of the input qubits adds a certain polynomial to the system, and one more polynomial arises from
the phase of a classical path through the circuit. Thereby, elements of the unitary matrix determined by the
quantum circuit under consideration can be computed by counting the number of common roots in $Z_2$
for the polynomial system associated with the circuit. Given a quantum circuit, the polynomial set
is uniquely constructed.

In this paper which is a substantially extended version of paper~\cite{GS05} we describe an algorithm for building
an arbitrary quantum circuit made of Hadamard and Toffoli gates and for constructing the associated polynomial system.
The algorithm implemented in the form of C\#~\cite{Csharp} program called QuPol (abbreviation of {\underline{Qu}antum}
{\underline{Pol}ynomials}). The program provides a user-friendly graphical interface for building
quantum circuits and generating the associated polynomial systems.

To build an arbitrary quantum circuit, it is represented as a rectangular table
with $N\times M$ cells, where $N$ the number of the input qubits, $M$ the number of the circuit
cascades plus one - for the input $N$-qubit vector. Each cell contains an elementary gate
taken from the elementary gate set. The last set of gates consists of wires (identities), arithmetic operations
(multiplication and addition) and the Hadamard operation. A user of QuPol can construct a
quantum circuit by selecting required elementary gates from the menu bar of the program and by
placing them in the appropriate cells. As a result, the constructed table provides a visual representation
of the circuit. A recursive evaluation procedure built-in QuPol
operates on the table column-by-column and generates $N$ polynomials (one for each row of the table)
as well as the phase polynomial in $n$ variables, where $n$ the number of Hadamard gates in the circuit.

To count the number of solutions in $Z_2$ for the output polynomial system, one can convert it
into an appropriate \Gr basis form which has the same number of solutions. Being invented 40 years ago~\cite{Buch65}, the \Gr bases has become
the most universal algorithmic method for analyzing and solving systems of polynomial equations~\cite{BW98}.
In particular, construction of the lexicographical \Gr basis substantially alleviates the problem of the root
finding for polynomial systems. To construct \Gr bases one can use, for example, efficient
involutive algorithms developed in~\cite{Gerdt}. Our QuPol program together with a \Gr basis software
provides a tool to analyze quantum computation by applying methods of modern computational commutative
algebra. We illustrate this tool by example from~\cite{Dawson}.

The structure of the paper is as follows. In Section~\ref{sec:quantumcircuits} we outline shortly
the circuit model of quantum computation. Section~\ref{sec:someoverpaths} presents the famous
Feynman's sum-over-paths method applied to quantum circuits. In Section~\ref{sec:algorithm} we
consider a circuit decomposition in terms of the elementary gates and show how to build an
arbitrary circuit composed from the Hadamard and Toffoli gates. These two gates form a universal gate
basis~\cite{Shi,Aharonov}. Some features of the algorithm implementation are briefly described.
Section~\ref{sec:matrixelements} demonstrates a simple example of handling the polynomials
associated with a quantum circuit by constructing their \Gr basis and computing the circuit matrix.
We conclude in Section~\ref{sec:conclusions}.

\section{QUANTUM CIRCUITS}
\label{sec:quantumcircuits}

To compute a reversible Boolean vector-function $f:\ZZ^n\to
\ZZ^n$, one applies an appropriate unitary transformation $U_f$
to an input state $\left| \mathbf{a} \right\rangle$ composed of
of $n$ qubits~\cite{Nielsen}:
  $$
  \left| \mathbf{b} \right\rangle  = U_f \left| \mathbf{a}
  \right\rangle, \qquad \left| \mathbf{a}
  \right\rangle ,\left| \mathbf{b} \right\rangle \in \CC^{2^n}\,.
  $$
The output state $\left| \mathbf{b} \right\rangle$ is generally not
required result of the computation until somebody observes it.
After that the output state becomes classical and can be used
anywhere.

Some elementary unitary transformations are called quantum gates.
A quantum gate acts only on a few qubits, on the remaining qubits it acts as
the identity. One can build a quantum circuit by appropriately
aligning quantum gates. In so doing, the unitary transformation of the circuit
is the composition of its elementary unitary transformations:
  \begin{equation}
  U_f  = U_m U_{m - 1}  \cdots U_2 U_1\,.
  \label{eq:ufdecomposition}
  \end{equation}

A quantum gate basis is a set of universal quantum gates, i.e. any
unitary transformation can be presented as a composition of the
gates of the basis. As well as in the classical case, there can be
different choice of the basis quantum gates~\cite{Nielsen}. For our work it is
convenient to choose the particular universal gate basis
consisting of Hadamard and Toffoli gates~\cite{Shi,Aharonov}.

The Hadamard gate is a one-qubit gate. It turns the computational
basis states into the equally weighted superpositions
  \begin{eqnarray*}
  && H:\left| 0 \right\rangle  \mapsto \frac{1}{{\sqrt 2 }}(\left| 0 \right\rangle  +
  \left| 1 \right\rangle )\,, \\
  && H:\left| 1 \right\rangle  \mapsto \frac{1}{{\sqrt 2 }}(\left| 0 \right\rangle  -
  \left| 1 \right\rangle ) \\
  \end{eqnarray*}
which differ in the phase factor at $\left| 1 \right\rangle$.

The Toffoli gate is a tree-qubit gate. The input bits $x$ and $y$
control the operation on bit $z$, and the Toffoli gate acts on
the computational basis states as~\cite{Dawson}
  $$
  \left( {x,y,z} \right) \mapsto \left( {x,y,z \oplus xy} \right)\,.
  $$

An action of a quantum circuit can be described by a square
unitary matrix whose entry $\left\langle \mathbf{b}
\right|U_f \left| \mathbf{a} \right\rangle$ yields the probability
amplitude for the transition from the initial quantum state $\left|
\mathbf{a} \right\rangle$ to the final quantum state $\left|
\mathbf{b} \right\rangle$. The matrix element is represented in
accordance to the gate decomposition~(\ref{eq:ufdecomposition}) of the circuit unitary
transformation and can be calculated as
sum over all the intermediate states $\mathbf{a}_i$, i = 1,2,
\ldots m - 1:
  $$
  \left\langle \mathbf{b} \right|U_f \left| \mathbf{a} \right\rangle
  = \sum\limits_{\mathbf{a}_i } {\left\langle \mathbf{b} \right|U_m
  \left| {\mathbf{a}_{m - 1} } \right\rangle } \cdots \left\langle
  {\mathbf{a}_1 } \right|U_1 \left| \mathbf{a} \right\rangle\,.
  $$

\section{SUM-OVER-PATHS FOR QUANTUM CIRCUITS}
\label{sec:someoverpaths}

To apply the famous quantum-mechanical Feynman's sum-over-paths approach to calculation of the matrix
elements for a quantum
circuit~\cite{Dawson}, we replace each quantum gate in the circuit under consideration by its
classical counterpart. The trick here is to determine the corresponding classical gate for the quantum
Hadamard gate since its action at any input value 0 or 1 must generate 0 or 1 with the equal probability.
To take this into account, the output of the classical Hadamard gate can be characterized by the path variable
$x\in \ZZ$~\cite{Dawson}. Its value determines
one of the two possible paths of computation. Thereby, the classical Hadamard gate as
$$
a_1  \mapsto x\,, \qquad a_i ,x \in \ZZ\,,
$$
and the classical Toffoli gate acts as
$$ \left( {a_1
,a_2 ,a_3 } \right) \mapsto \left( {a_1 ,a_2 ,a_3 \oplus a_1 a_2 }
\right)\,
$$
where $\oplus$ denotes addition modulo 2.

\begin{figure}[h!]
   \begin{center}
   \begin{tabular}{c}
   \includegraphics[width=6cm]{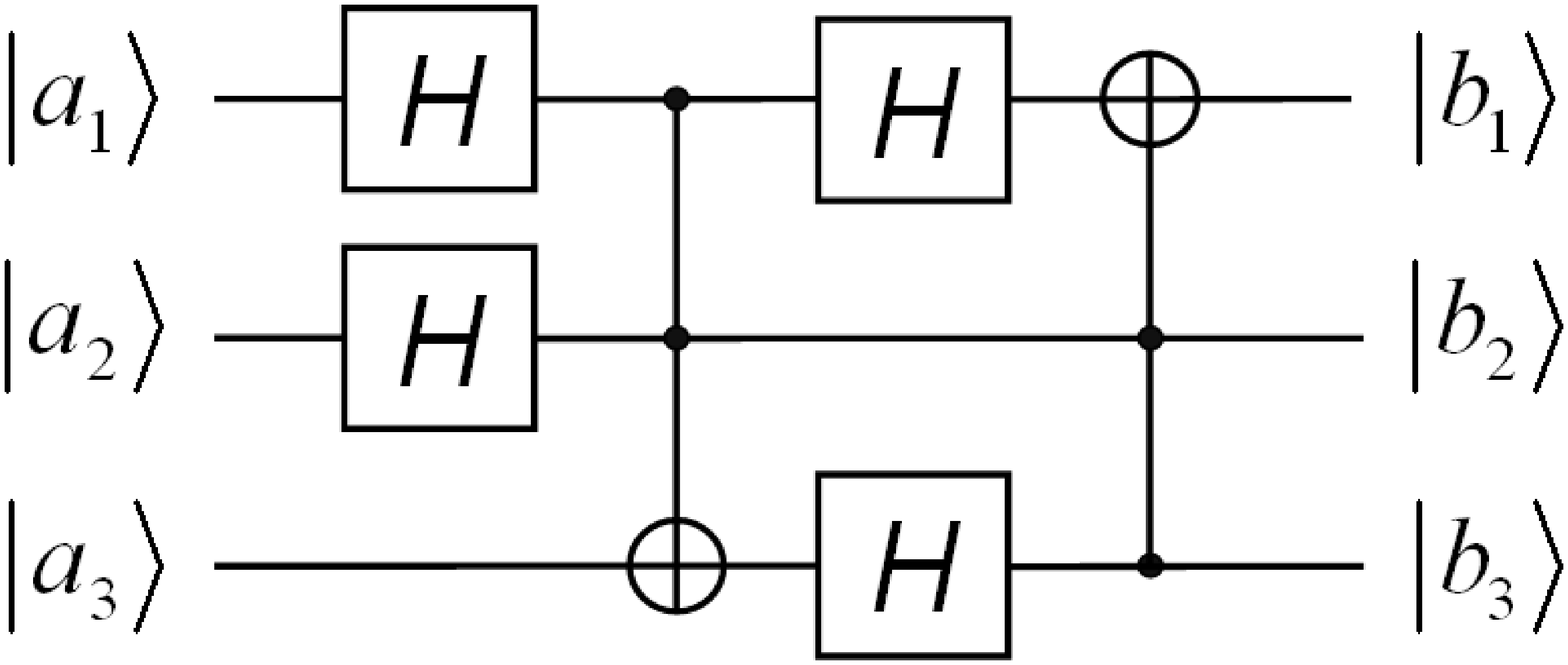}
   \raisebox{1\height}{\includegraphics[width=.7cm]{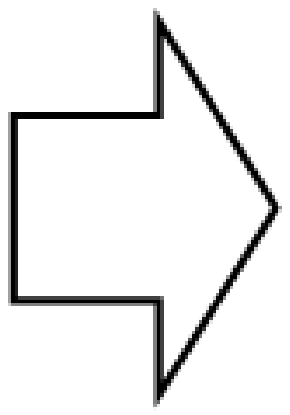}}
   \includegraphics[width=9.5cm]{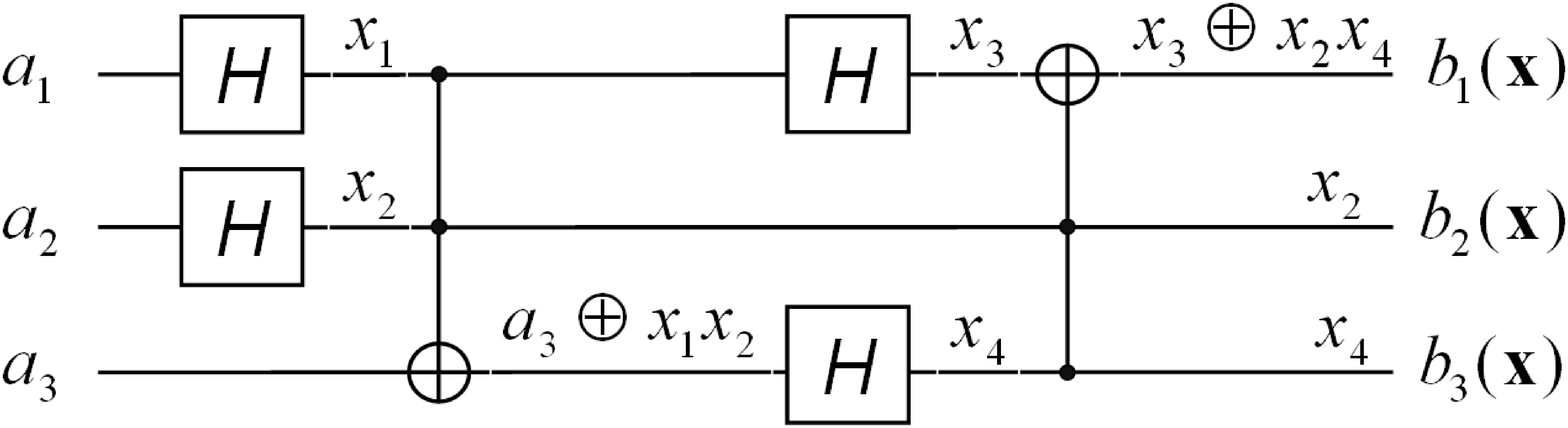}
   \end{tabular}
   \end{center}
   \caption[From quantum to classical circuit]{
   \label{fig:quantum2classical}
   From quantum to classical circuit}
   \end{figure}

Fig.~\ref{fig:quantum2classical} shows an example of quantum
circuit (taken from~\cite{Dawson}) and its classical
counterpart. The path variables $x_i$ comprise the (vector)
path $\mathbf{x} = (x_1 ,x_2 ,x_3 ,x_4 )^T  \in \ZZ^4$.

\begin{figure}[h!]
   \begin{center}
   \begin{tabular}{c}
   \includegraphics[width=9.5cm]{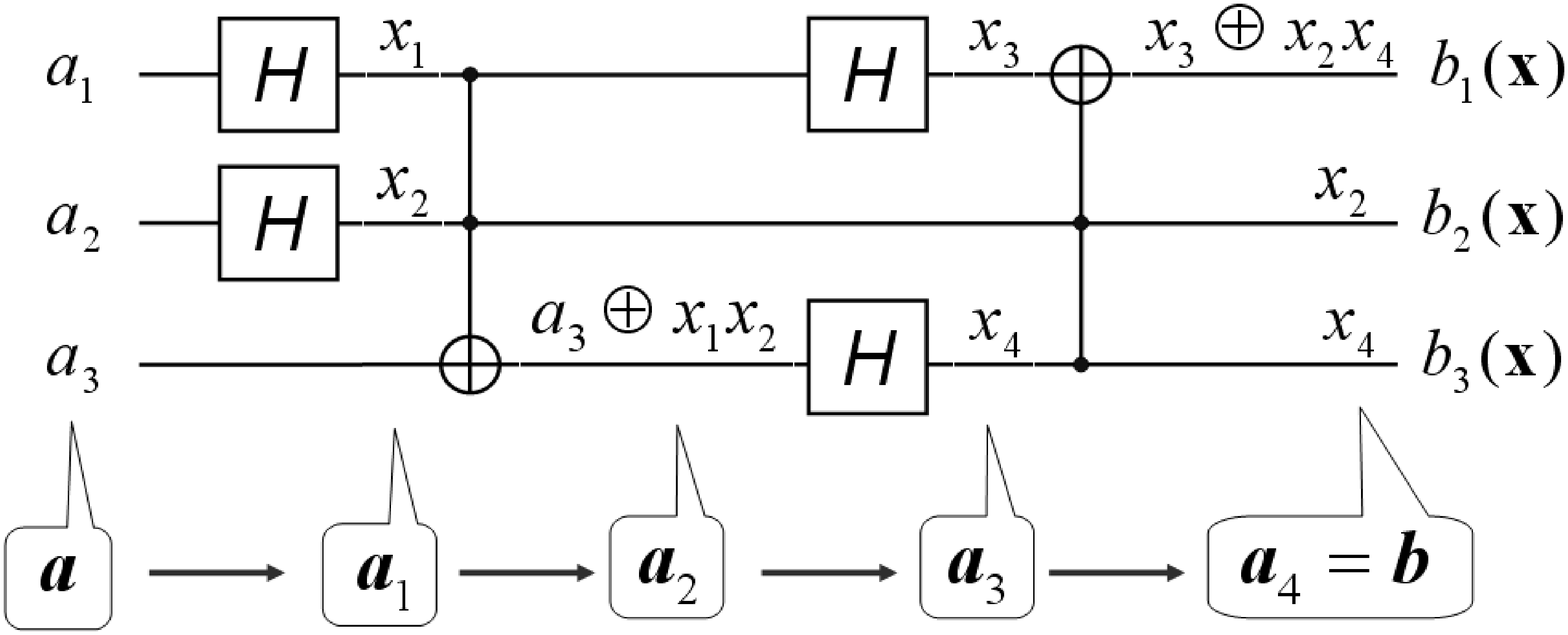}
   \end{tabular}
   \end{center}
   \caption[Admissible classical paths]{
   \label{fig:paths}
   Admissible classical paths}
   \end{figure}

A classical path is defined by a sequence of classical bit strings $\mathbf{a},\mathbf{a_1} ,\mathbf{a_2} ,
\ldots ,\mathbf{a_m} = \mathbf{b}$ obtained from application of the classical gates. Each
set of values of the path variables $x_i$ gives a sequence of classical bit strings which
is called an admissible classical path (Fig.~\ref{fig:paths}).
Each admissible classical path has a phase which is determined by
the Hadamard gates applied~\cite{Dawson}. The phase is changed only when the
input and output of the Hadamard gate are simultaneously equal to
1, and this derives the formula
$$
\varphi (\mathbf{x}) = \sum\limits_{{\rm{Hadamard\ gates}}} {input
\bullet output}\,.
$$
with the sum evaluated in $\ZZ$. As to Toffoli gates, they do not change the phase.

\noindent
In our example the phase of the path $\mathbf{x}$ is
$$
\varphi (\mathbf{x}) = a_1 x_1  \oplus a_2 x_2  \oplus x_1 x_3
\oplus x_4 (a_3  \oplus x_1 x_2 )\,.
$$

The matrix element of a quantum circuit is the sum over all
the allowed paths from the classical state $\mathbf{a}$ to
$\mathbf{b}$
$$
\left\langle \mathbf{b} \right|U_f \left| \mathbf{a} \right\rangle
= \frac{1}{{\sqrt {2^h }
}}\sum\limits_{\mathbf{x}:\mathbf{b}\left( \mathbf{x} \right) \mathbf{b}} {\left( { - 1} \right)} ^{\varphi \left( \mathbf{x}
\right)}\,,
$$
where $h$ is the number of Hadamard gates. The terms in the sum
have the same absolute value but may vary in sign.

Let $N_0$ be the number of positive terms in the sum and $N_1$ be the
number of negative terms:
\begin{eqnarray}
&& N_0  =\ \mid {\left\{\ {\mathbf{x}\mid \mathbf{b}(\mathbf{x}) = \mathbf{b}\quad \wedge \quad \varphi
 (\mathbf{x}) = 0\ }
\right\}} \mid \,,  \label{N0} \\
&& N_1  =\ \mid {\left\{\ {\mathbf{x}\mid \mathbf{b}(\mathbf{x}) = \mathbf{b}\quad \wedge \quad \varphi
 (\mathbf{x}) = 1\ }
\right\}} \mid\,. \label{N1}
\end{eqnarray}
Hence, $N_0$ and $N_1$ count the number of solutions for the indicated systems of $n+1$ polynomials
in $h$ variables over $\ZZ$. Then the matrix element may be
written as the difference
\begin{equation}
\left\langle \mathbf{b} \right|U_f \left| \mathbf{a} \right\rangle
= \frac{1}{{\sqrt {2^h } }}\left( {N_0  - N_1 } \right)\,. \label{matrix}
\end{equation}

\section{ALGORITHM AND PROGRAM}
\label{sec:algorithm}

\begin{figure}[h!]
   \begin{center}
   \begin{tabular}{c}
   \includegraphics[width=4.2cm]{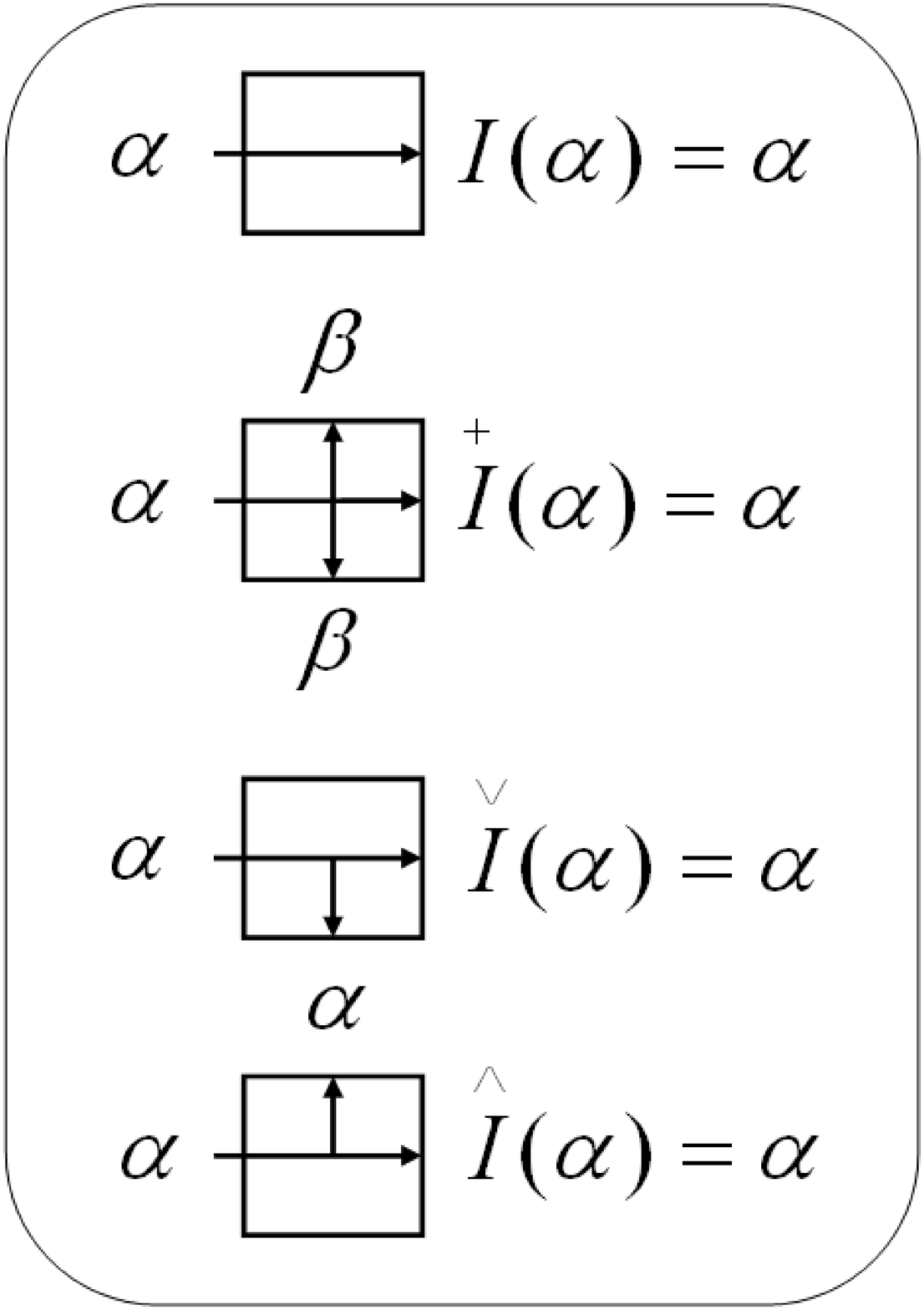}
   \includegraphics[width=4.2cm]{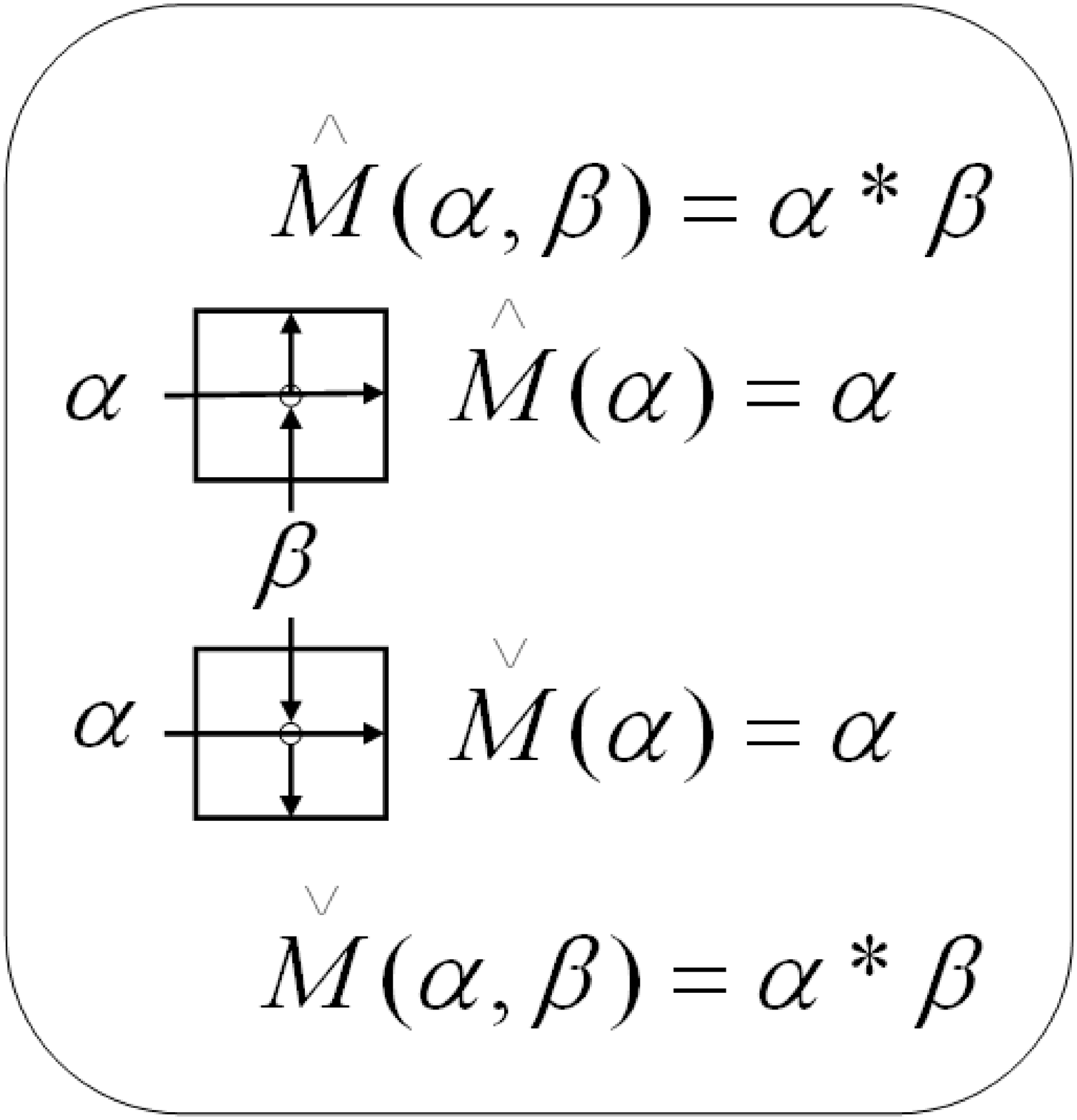}
   \includegraphics[width=5.2cm]{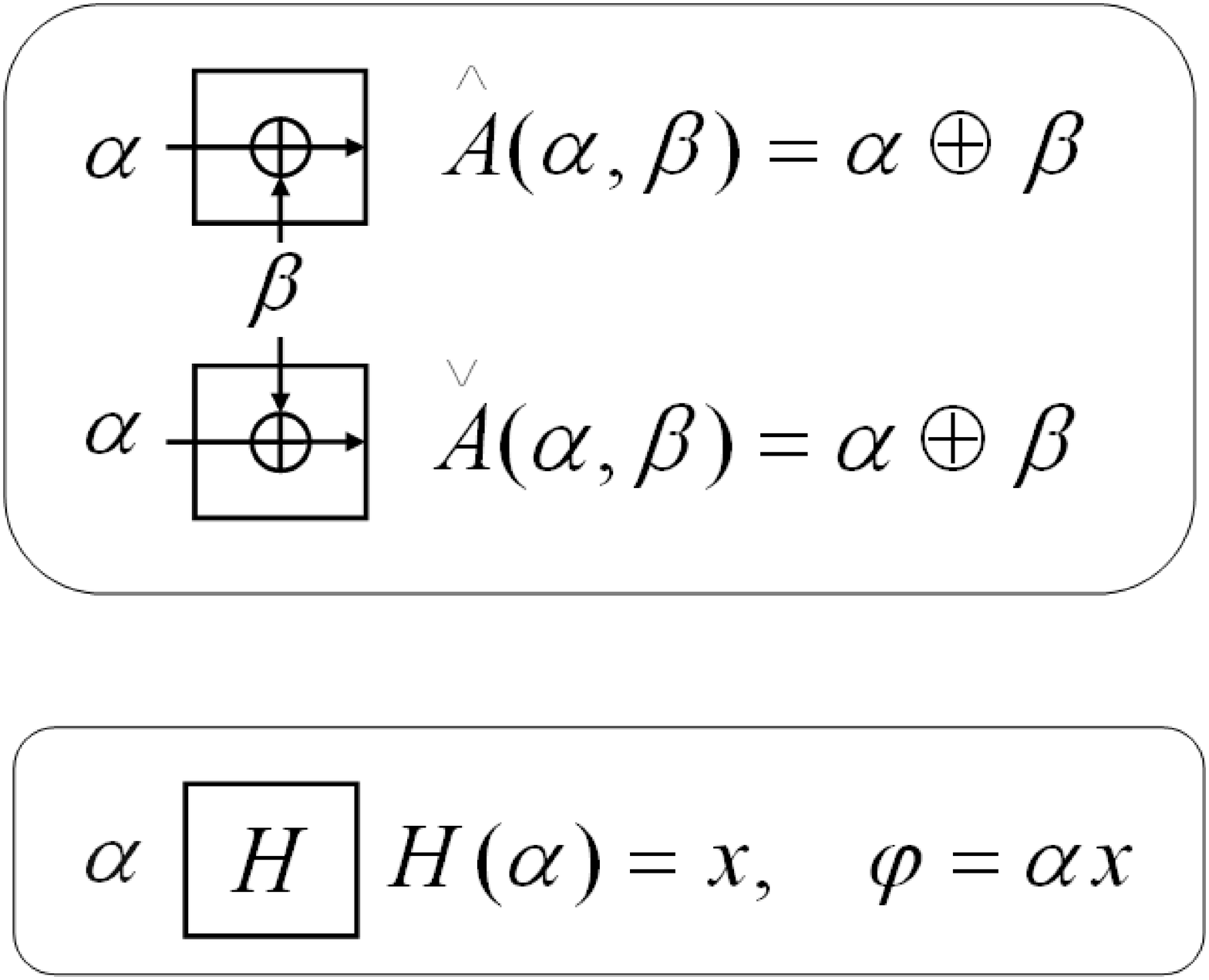}
   \end{tabular}
   \end{center}
   \caption[Elementary gates]
   {\label{fig:elelementarygates}
   Elementary gates}
   \end{figure}

\begin{figure}[h!]
   \begin{center}
   \begin{tabular}{c}
   \includegraphics[width=7cm]{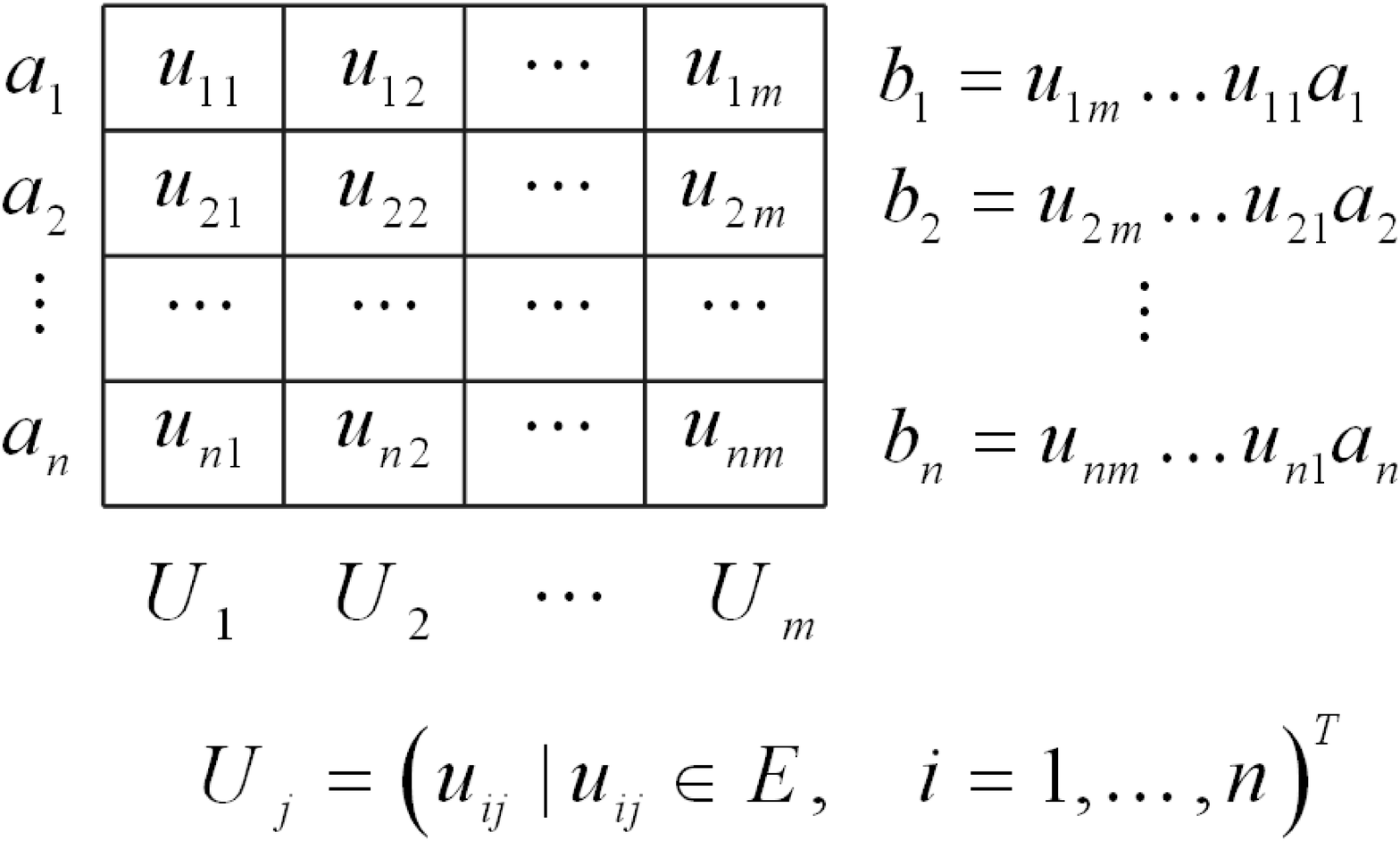}
   \hspace{3cm}
   \includegraphics[width=5cm]{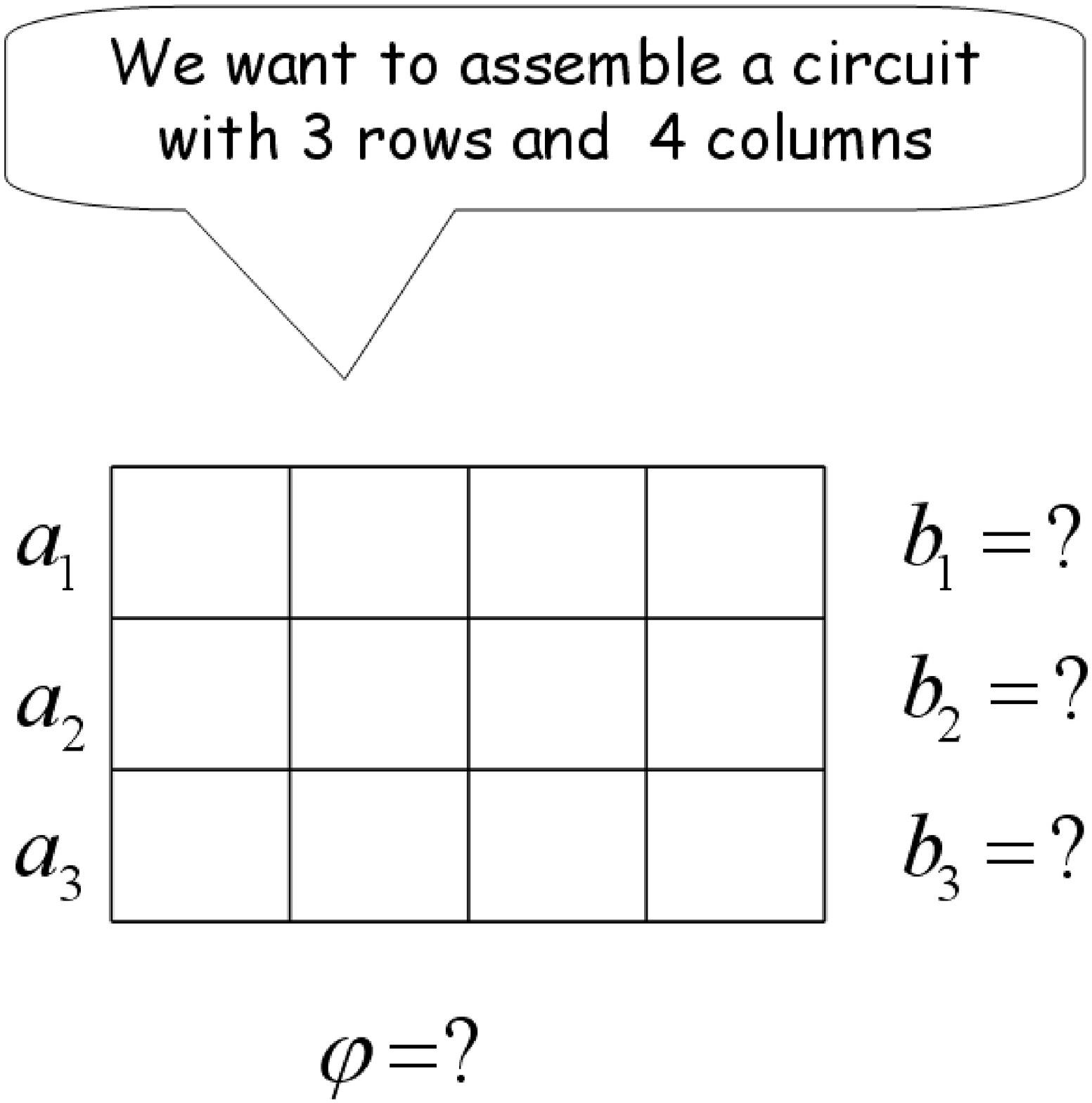}
   \end{tabular}
   \end{center}
   \caption[decomposition]
   { \label{fig:decomposition}
   Decomposition into elementary gates (left) and building a sample
   circuit, step 0 (right)}
   \end{figure}

For building arbitrary quantum circuits from Hadamard
and Toffoli gates, we use the set of elementary gates
  \begin{equation}
  E = \{ I,\mathop {I,}\limits^ +  \mathop {I,}\limits^ \wedge
  \mathop I\limits^ \vee  ,\mathop M\limits^ \wedge  ,\mathop
  M\limits^ \vee  ,\mathop A\limits^ \wedge  ,\mathop A\limits^ \vee
  ,H\} \label{elementary gates}
  \end{equation}
shown on Fig.~\ref{fig:elelementarygates}.  Furthermore, we represent a circuit as a rectangular table
(Fig.~\ref{fig:decomposition}, left).

Each cell in the table contains an elementary gate so that the output for each row is determined by
the composition of the elementary gates in the row (taking into account recursive dependences).
Hence, each elementary unitary transformation $U_j$ is represented by an $n$-tuple of elementary
gates.

Fig.~\ref{fig:elelementarygates} shows action of the elementary
gates from the set (\ref{elementary gates}): the identities, the
multiplications, the additions modulo 2, and the classical
Hadamard gate. The identity $I$ just reproduces its input. The
identity-cross $\overset{+}{I}$ reproduces also its vertical input
from the top elementary gate to the bottom one and vice-versa.
Every identity-down $\overset{\vee}{I}$ and identity-up
$\overset{\wedge}{I}$ have two outputs -- horizontal and vertical.
The multiplication-up $\overset{\wedge}{M}$ and
multiplication-down $\overset{\vee}{M}$ perform multiplication of
their horizontal and the corresponding vertical inputs. The addition-up $\overset{\wedge}{A}$ and
addition-down $\overset{\vee}{A}$ are applied similarly. Each Hadamard gate outputs an
independent path variable irrespective of its input and can give a
nonzero contribution to the phase.
    \begin{figure}
   \begin{center}
   \begin{tabular}{c}
   \includegraphics[width=5.7cm]{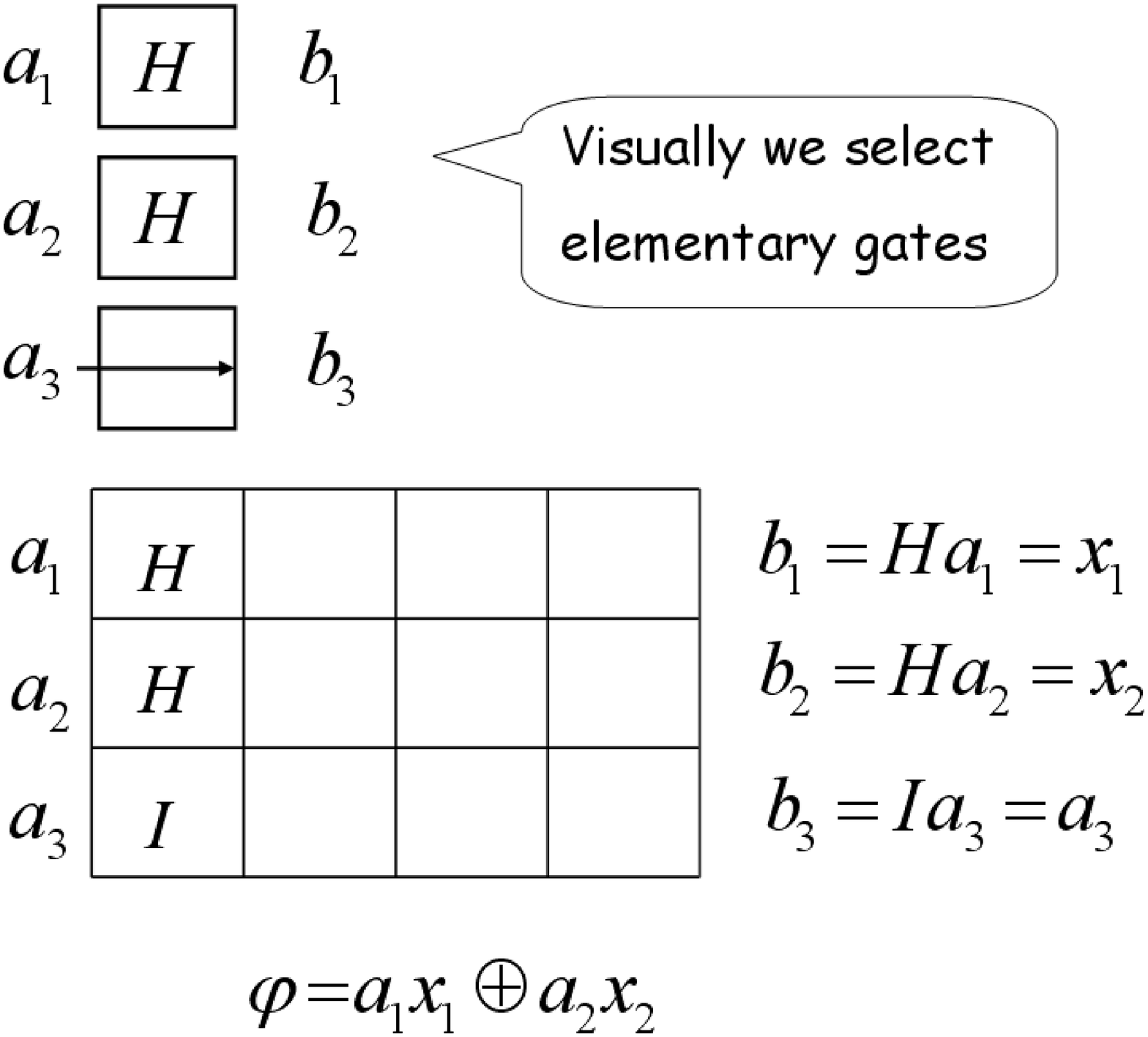}
   \hspace{3cm}
   \includegraphics[width=6.5cm]{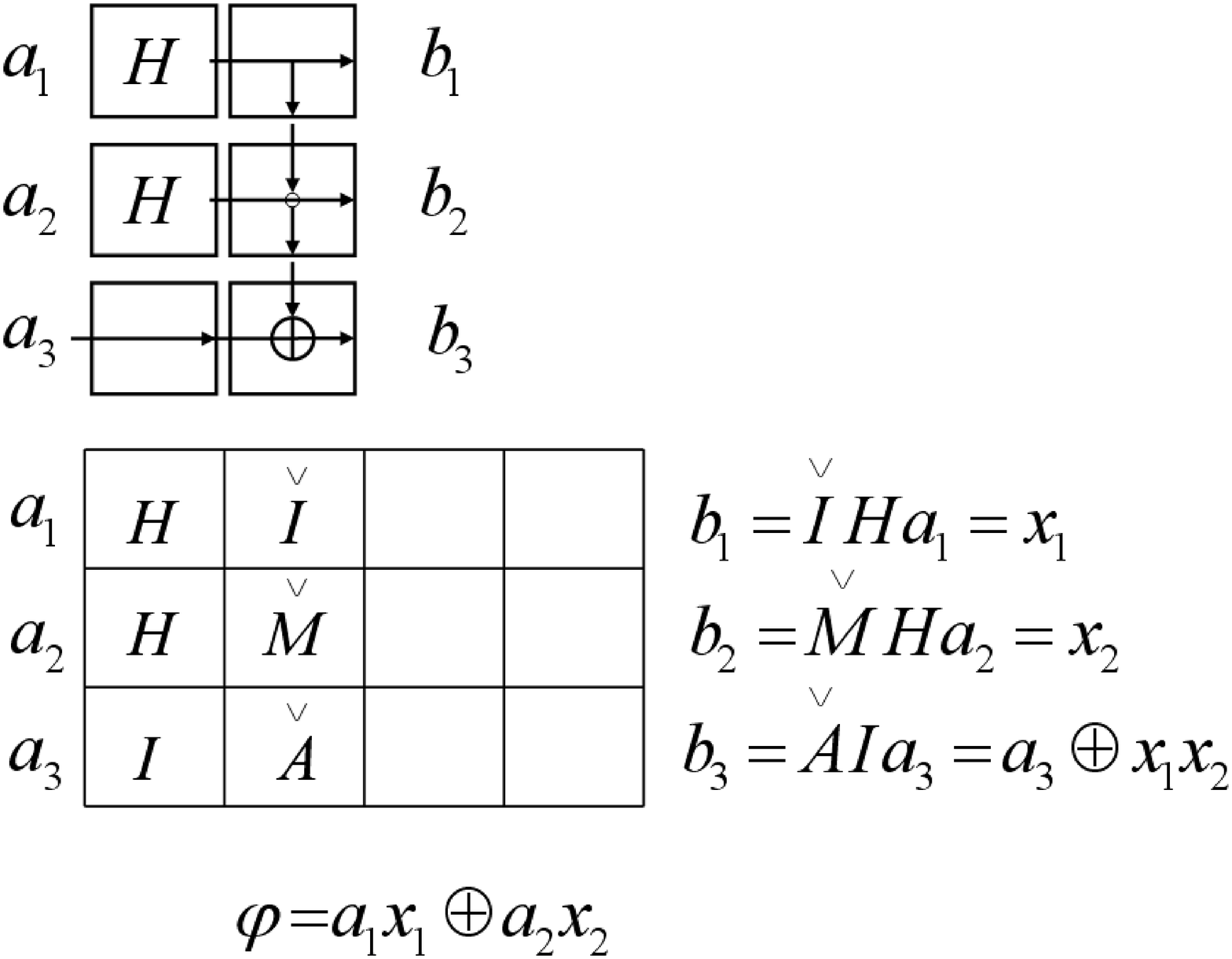}
   \end{tabular}
   \end{center}
   \caption[steps12]
   { \label{fig:steps12}
   Building circuit, steps 1 (left) and 2 (right)}
   \end{figure}
   \begin{figure}
   \begin{center}
   \begin{tabular}{c}
   \includegraphics[width=6cm]{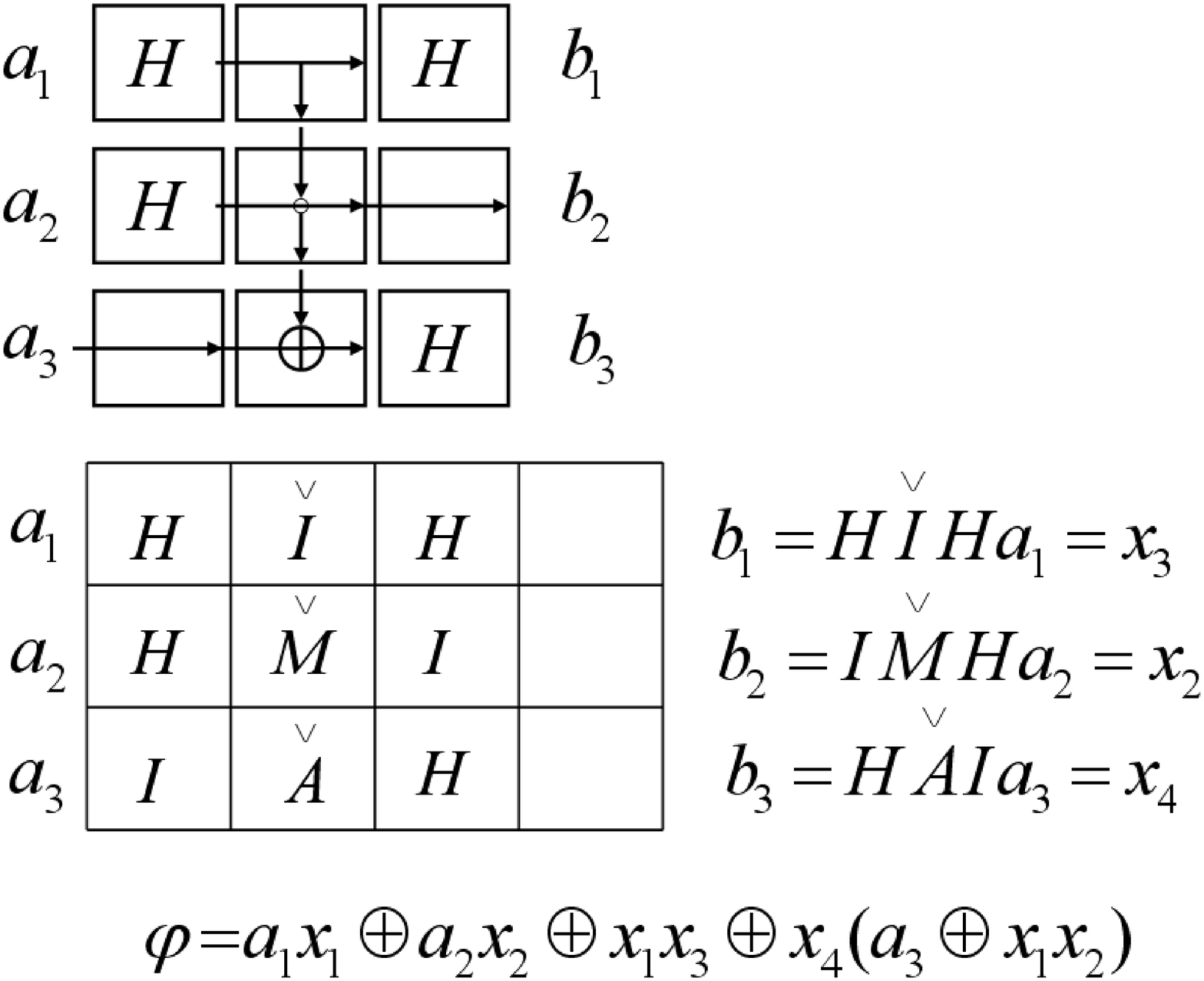}
   \hspace{2cm}
   \includegraphics[width=7cm]{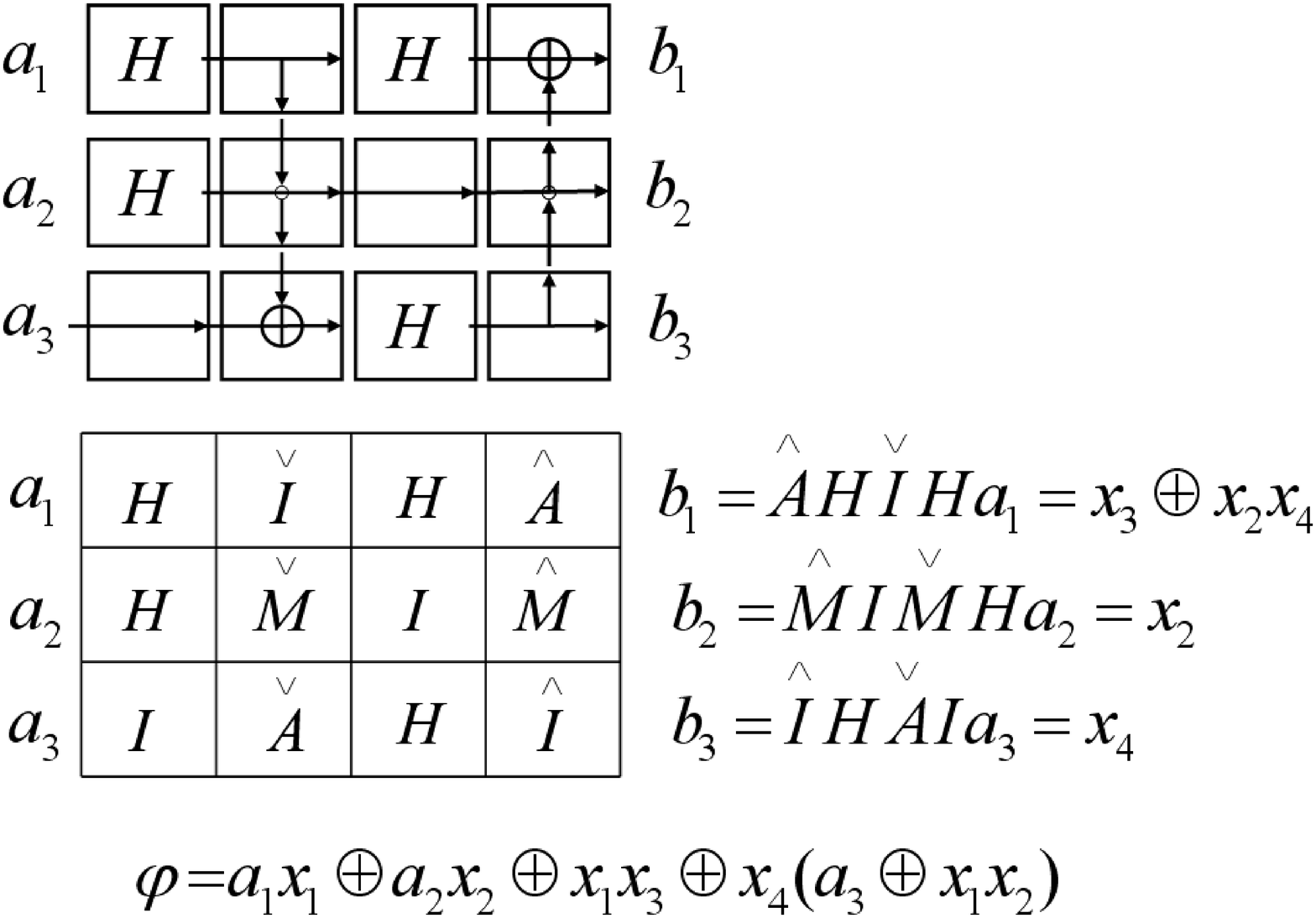}
   \end{tabular}
   \end{center}
   \caption[steps34]
   { \label{fig:steps34}
   Building circuit, steps 3 (left) and 4 (right)}
   \end{figure}

To build a circuit, we define, first, an empty table of the required
size (Fig.~\ref{fig:decomposition}, right). At this initialization step the
output qubit values and phase are not determined. As the second step we place the required
elementary gates in appropriate cells of the first column and
construct the circuit polynomials (Fig.~\ref{fig:steps12}, left).
Now the output is obtained by applying the inserted elementary gates to
the input. The phase is also calculated. Then we proceed the same
way with the second column (Fig.~\ref{fig:steps12}, right), then with the third and forth columnns
(Fig.~\ref{fig:steps34}).

A circuit is represented in our program QuPol as two $2d$-arrays: one
for the elementary gates and another for the related polynomials. The
phase polynomial is separately represented. The following piece of
code demonstrates construction of the circuit polynomials

\begin{verbatim}
  for each Column in Table of Gates
    for each Gate in Column {
      construct Gate Polynomial;
      if Gate id Hadamard
        reconstruct Phase Polynomial; }
\end{verbatim}

It should be noticed that Figs.~\ref{fig:steps12} and \ref{fig:steps34} show a shortened version of the
elementary gates composition. Actually, the final polynomials look like:
\begin{eqnarray*}
&& b_1  = \mathop A\limits^ \wedge  (H\mathop I\limits^ \vee  H{\kern 1pt} a_1 ,\mathop M\limits^ \wedge  (I\mathop M\limits^ \vee  (H{\kern 1pt} a_{2,} \mathop I\limits^ \vee  H{\kern 1pt} a_1 ),\mathop I\limits^ \wedge  H\mathop A\limits^ \vee  (I{\kern 1pt} a_3 ,\mathop M\limits^ \vee  ({\kern 1pt} Ha_{2,} \mathop I\limits^ \vee  H{\kern 1pt} a_1 )))) = x_3  \oplus x_2 x_4 \,, \\
&& b_2  = \mathop M\limits^ \wedge  (I\mathop M\limits^ \vee  (H{\kern 1pt} a_2 ,\mathop I\limits^ \vee  H{\kern 1pt} a_1 ),\mathop I\limits^ \wedge  H\mathop A\limits^ \vee  (I{\kern 1pt} a_3 ,\mathop M\limits^ \vee  ({\kern 1pt} Ha_{2,} \mathop I\limits^ \vee  H{\kern 1pt} a_1 ))) = x_2 \,, \\
&& b_3  = \mathop I\limits^ \wedge  H\mathop A\limits^ \vee  (I{\kern 1pt} a_3 ,\mathop M\limits^ \vee  (H{\kern 1pt} a_{2,} \mathop I\limits^ \vee  H{\kern 1pt} a_1 )) = x_4 \,.
 \end{eqnarray*}

The method for constructing a gate polynomial is essentially recursive because
of the need to go up or down for some gates.

Any circuit is saved in the program as two files. One file is binary and contains
the circuit itself. Another file has a text format. It contains
the circuit polynomials in a symbolic form. The program allows one to
save polynomials in several formats suitable as inputs for
computer algebra systems such as Maple or Mathematica  for
the further processing. It is also possible to load the saved circuit back into
the computer memory for its modification.

\begin{figure}[h!]
   \begin{center}
   \begin{tabular}{c}
   \includegraphics[width=9cm]{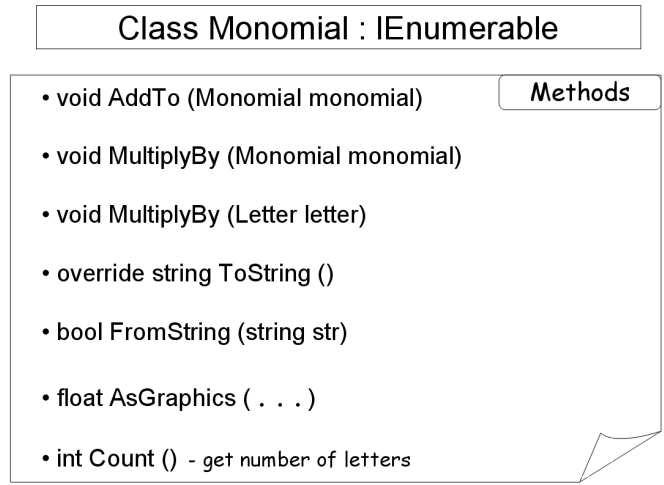}
   \end{tabular}
   \end{center}
   \caption[Class Monomial]
   { \label{fig:classmonomial}
   Class Monomial}
   \end{figure}

\begin{figure}[h!]
   \begin{center}
   \begin{tabular}{c}
   \includegraphics[width=9cm]{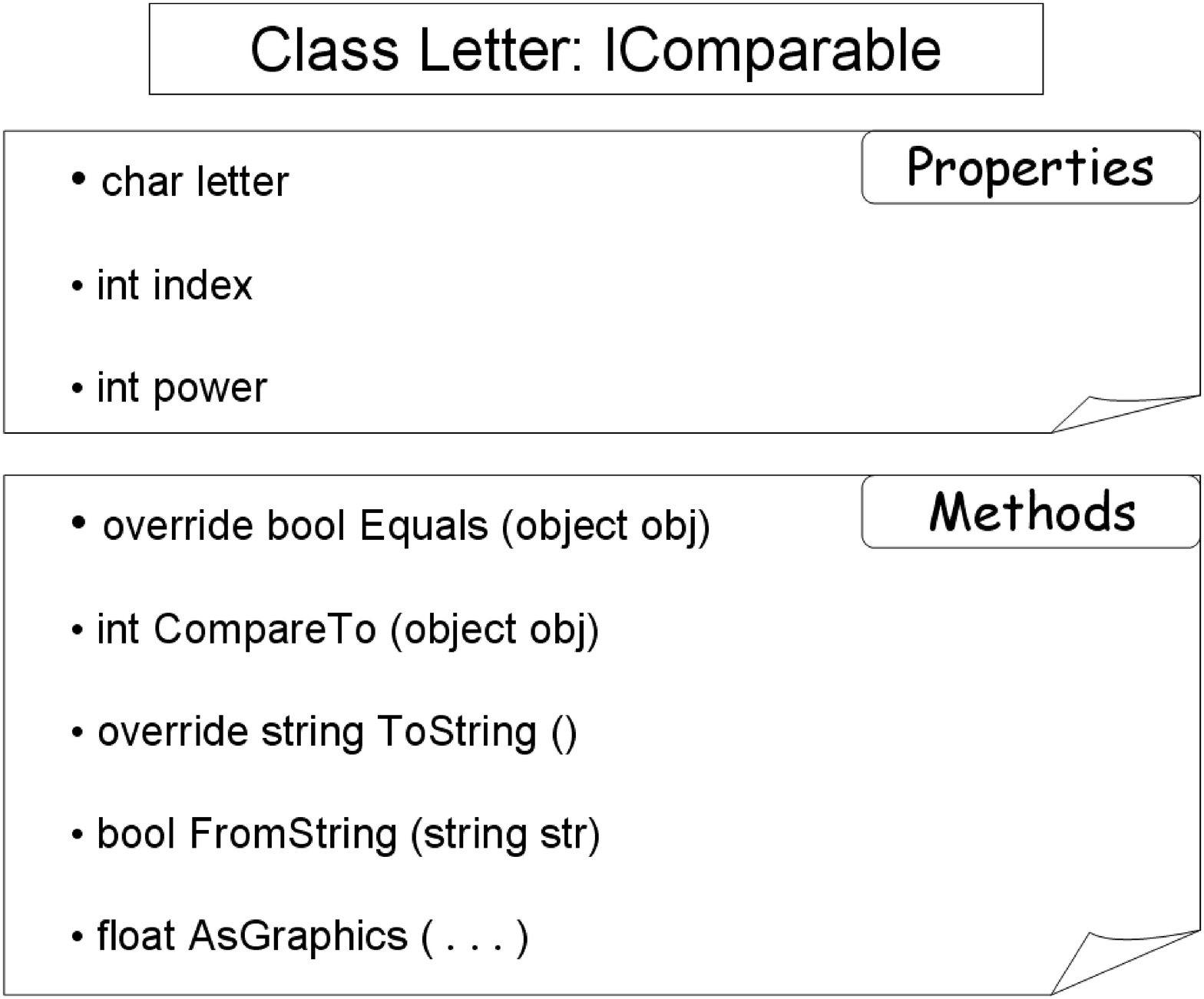}
   \end{tabular}
   \end{center}
   \caption[Class Letter]
   { \label{fig:classletter}
   Class Letter}
   \end{figure}

The part of our code for the name space {\bf
Polynomial\_Modulo\_2} written in C\# can be also
   \begin{figure}[h!]
   \begin{center}
   \begin{tabular}{c}
   \includegraphics[width=9cm]{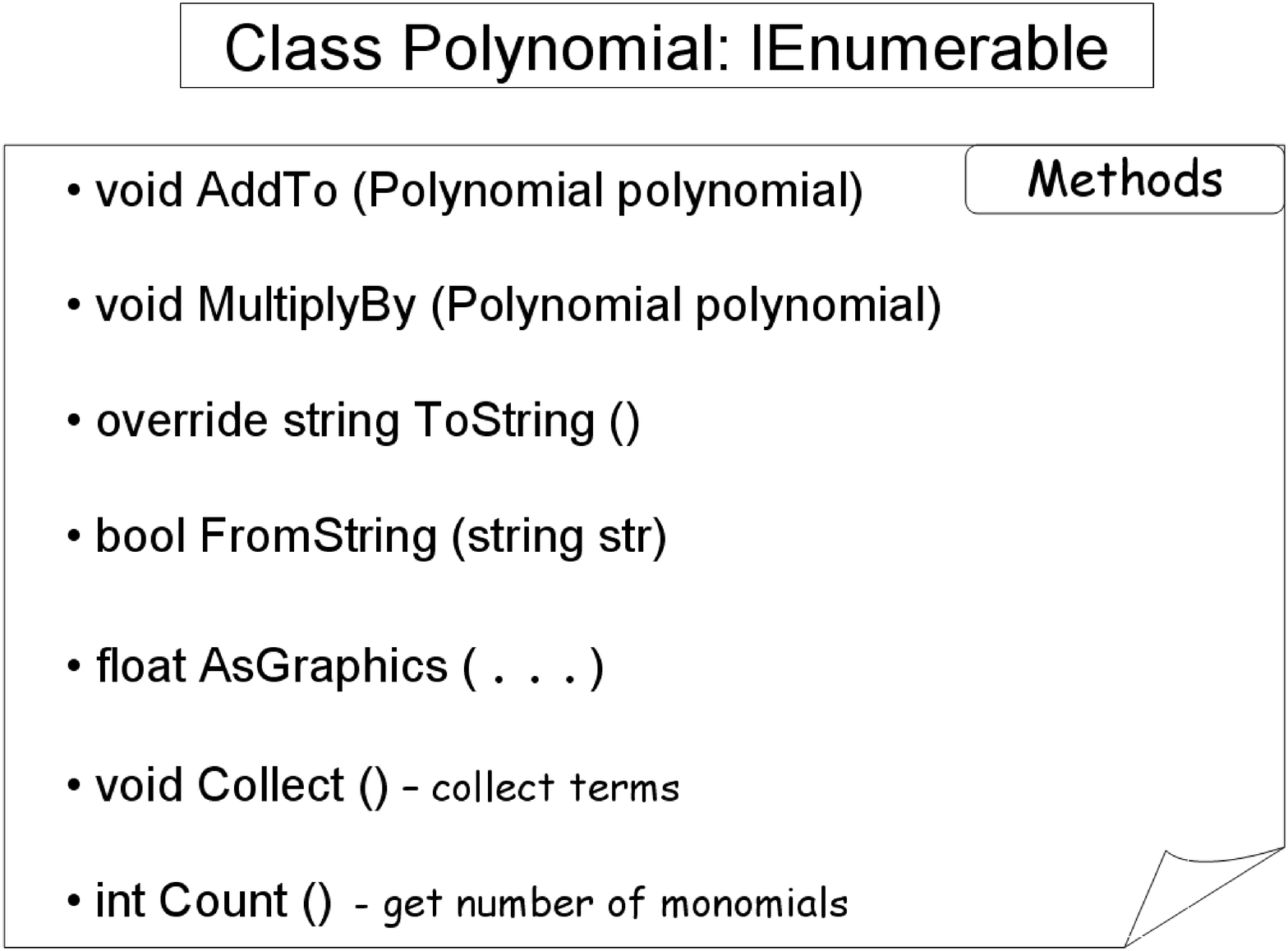}
   \end{tabular}
   \end{center}
   \caption[Class Polynomial]
   { \label{fig:classpolynomial}
   Class Polynomial}
   \end{figure}
used independently on our program. This name space contains
a set of classes for handling polynomials over the field $\ZZ$.
Class Polynomial (Fig.~\ref{fig:classpolynomial}) is a list of
monomials, class Monomial (Fig.~\ref{fig:classmonomial}) is a
list of letters, class Letter (Fig.~\ref{fig:classletter}) is an
indexed letter provided with a positive integer superscript (power
degree).

\section{CALCULATING CIRCUIT MATRIX ELEMENTS}
\label{sec:matrixelements}

A system generated by the program is a finite set $ F \subset \,R
$ of polynomials in the ring
  \begin{equation}
  R= \ZZ [a_i ,b_j ][x_1 ,...,x_h ]\,, \quad
  a_i ,b_j  \in \ZZ\,,\quad i,j = 1,\ldots,n \label{ring}
  \end{equation}
with $h$ polynomial variables and $2n$ binary coefficients. Let $N_0$ and $N_1$ denote, as
in~(\ref{N0}-\ref{N1}),
the number of roots in  $\ZZ$ of the polynomial sets $F_0$ and $F_1$, respectively, where
  \begin{equation}
  F_0  = \{ f_{} ,...,f_k ,\varphi \,\}\,, \quad F_1  = \{ f_{}
  ,...,f_k ,\varphi  \oplus 1\,\} \,. \label{circuit systems}
  \end{equation}

Then the circuit matrix is given as
  $$
  \left\langle {\mathbf{b} } \right|\,U\,\left| {\,\mathbf{a} }
  \right\rangle = \frac{1}{{\sqrt {2^{\,h} } }}\,\,(\,N_0  - N_1\,)\,,
  $$
where $h$ is the number of Hadamard gates in the circuit.

To compute $N_0$ and $N_1$ one can convert $F_0$ and $F_1$ into
an appropriate triangular form~\cite{Wang} providing elimination of the path variables $x_1 ,...,x_h$.
One of such triangular forms is the pure lexicographical Gr\"{o}bner basis that can be computed,
for instance, by means of the Buchberger algorithm~\cite{Buch65,BW98}, by the Faug\`ere
algirthms~\cite{F4,F5}  or by our involutive algorithm~\cite{Gerdt}.

For the circuit shown on Fig.~\ref{fig:quantum2classical} we obtain
the following polynomial set:
  \begin{equation}
  \begin{array}{l}
  f_1  = x_2 x_4 \, \oplus x_3  \oplus b_1\,,  \\
  f_2  = x_2  \oplus b_2 \,, \\
  f_3  = x_4  \oplus b_3  \,,\\
  \varphi  = \,x_1 x_2  \oplus x_1 x_3  \oplus a_1 x_1  \oplus a_2 x_2  \oplus a_3 x_4 \,. \\
  \end{array} \label{pol_system}
 \end{equation}

The lexicographical Gr\"{o}bner bases for polynomial systems $F_0$ and $F_1$
and for the ordering $ x_1 \succ
x_2 \succ x_3 \succ x_4 \,\,$ on the variables are given by

\begin{equation}
  G_0:\ \left\{
  \begin{array}{l}
  g_1  = (a_1  \oplus b_1 )x_1 \, \oplus a_2 b_2  \oplus a_3 b_3\,, \\
  g_2  = x_2  \oplus b_2 \,, \\
  g_3  = \,\,x_3  \oplus b_1  \oplus b_2 b_3\,,  \\
  g_3  = x_4  \oplus b_3 \,, \\
  \end{array}
  \right.
  \quad
  G_1:\ \left\{
  \begin{array}{l}
  g_1  = (a_1  \oplus b_1 )x_1 \, \oplus a_2 b_2  \oplus a_3 b_3 \oplus 1\,,  \\
  g_2  = x_2  \oplus b_2 \,, \\
  g_3  = \,\,x_3  \oplus b_1  \oplus b_2 b_3 \,, \\
  g_3  = x_4  \oplus b_3 \,. \\
  \end{array}
  \right.  \label{GB_systems}
\end{equation}

These lexicographical Gr\"{o}bner bases immediately yield
the following conditions on the parameters:
\begin{eqnarray}
  && G_0:\quad a_1  \oplus b_1  = a_2 b_2  \oplus a_3 b_3  = 0\,, \label{cond1}\\
  && G_1:\quad a_1  \oplus b_1  = 0 \ \wedge \ a_2 b_2  \oplus a_3 b_3  =1\,. \label{cond2}
\end{eqnarray}

It is immediately follows that if conditions~(\ref{cond1}) are satisfied then
the polynomial system $G_0$ (resp. $F_0$) has two common roots in $\ZZ$ and  $G_1$ (resp. $F_1$)
has no common roots, and, vise-versa, if conditions~(\ref{cond1}) are satisfied then $G_0$ has no roots
and $G_1$ has two roots. In all other cases there is one root of $G_0$ and one root
of $G_1$.

In that way, the $8\times 8$ matrix for the circuit of Fig.\ref{fig:quantum2classical} is easily determined
by formulae (\ref{matrix}) with $N_0$ and $N_1$ defined from systems~(\ref{GB_systems}). Table~\ref{table}
explicitly shows entries of this matrix when its rows and columns are indexed by the values of the input bits
$a_1,a_2,a_3$ and
output bits $b_1,b_2,b_3$, respectively.

\begin{table}[h]
\begin{center}
\begin{tabular} {|r|c|c|c|c|c|c|c|c|} \hline
     & & & & & & & & \\[-0.3cm]
    $b_1$& 0 & 0 & 0 & 0 & 1 & 1 & 1 & 1  \\
    $b_2$& 0 & 0 & 1 & 1 & 0 & 0 & 1 & 1  \\
    $b_3$& 0 & 1 & 0 & 1 & 0 & 1 & 0 & 1  \\[-0.2cm]
$a_1$\ $a_2$\ $a_3$    \, \ \,&  &  &  &  & &  &  &   \\
 \hline
0\ \ 0\,\,\ 0\ \ \ \ \ \ & 1/2 & 1/2 & 1/2 & 1/2 & 0 & 0 & 0 & 0 \\ \hline
0\ \ 0\,\,\ 1\ \ \ \ \ \ & 1/2 & -1/2 & 1/2 & -1/2 & 0 & 0 & 0 & 0 \\ \hline
0\ \ 1\,\,\ 0\ \ \ \ \ \ & 1/2 & 1/2 & -1/2 & -1/2 & 0 & 0 & 0 & 0 \\ \hline
0\ \ 1\,\,\ 1\ \ \ \ \ \ & 0 & -1/2 & -1/2 & 1/2 & 0 & 0 & -1/2 & 0 \\ \hline
1\ \ 0\,\,\ 0\ \ \ \ \ \ & 0 & 0 & 0 & 0 & 1/2 & 1/2 & 1/2 & 1/2 \\ \hline
1\ \ 0\,\,\ 1\ \ \ \ \ \ & 0 & 0 & 0 & 0 & 1/2 & -1/2 & 1/2 & -1/2 \\ \hline
1\ \ 1\,\,\ 0\ \ \ \ \ \ & 0 & 0 & 0 & 0 & 1/2 & 1/2 & -1/2 & -1/2 \\ \hline
1\ \ 1\,\,\ 1\ \ \ \ \ \ & 0 & 0 & 0 & 0 & 1/2 & -1/2 & -1/2 & 1/2 \\
\hline
\end{tabular}
\end{center}
\caption{The circuit matrix for polynomial system~(\ref{pol_system})}
\label{table}
\end{table}

\section{CONCLUSION}
\label{sec:conclusions}

We presented an algorithm for building an arbitrary quantum circuit composed of Hadamard and
Toffoli gates and for constructing the corresponding polynomial equation systems over $\ZZ$. The number of
common roots $\ZZ$ of polynomials in the system uniquely determines the circuit matrix. The
algorithm has been implemented in as a C\#-program QuPol.

To count the number of polynomial roots one can use the universal algorithmic Gr\"{o}bner basis approach.
In the framework of this approach the polynomial system under consideration is converted
into a triangular form that is convenient for computing the number of roots.

General theoretical complexity bound for computing \Gr bases (see, for instance,~\cite{Gathen}) is
double exponential in the number of polynomial variables $x_i$
(the number of path variables, i.e. Hadamard gates, in our case). However, the due to the fact that all the path
variables $x_i$,
the input and output parameters $a_j,b_k$  as well as the numerical coefficients of the polynomials
are elements in the finite field $\ZZ$  the \Gr basis computation is sharply simplified. Thus, based on the
algorithm in~\cite{F5}, the analysis of costs for computing \Gr bases for polynomials over $Z_2$, arising in
cryptography, revealed~\cite{Inria} only a single exponential complexity in the number of equations.

Therefore, the above presented algorithm together with an appropriate software for computing  a \Gr
basis and for counting its number of common roots in $\ZZ$ provides a tool
for simulating quantum logical circuits and for estimation of their computational power. For this
purpose we are planning to create a special module in the open source software GINV~\cite{ginv}.
At present, GINV contains implementation in C++ of our involutive algorithm~\cite{Gerdt} for
computing \Gr bases over the field of rational numbers and over the field ${\mathbb{Z}_p}$ where $p$ is a prime
number. By this reason, the current version of GINV can be used for computing \Gr bases for polynomial
systems~(\ref{circuit systems}) only if one considers $2n$ parameters $a_i,b_j$ in~(\ref{ring}) as
variables. In this case, to compute a triangular \Gr basis form of the circuit polynomial, one should use
an elimination term order for those parameters such that any term containing path variables $x_k$ is
higher than that any term containing parameters only.

Such computational scheme, however, increases the number of polynomial (path) variables by $2n$ what
may lead to exponential slowing down, as was pointed to above. That is why we expect that
an adaptation of the built-in algorithms and data structures to the polynomial
ring~(\ref{ring}) with parametric coefficients will substantially
increase computational efficiency of GINV in its application to analysis and
simulation of quantum circuits.

Recently another algorithmic approach was suggested~\cite{Kornyak} to work with polynomial systems
over finite fields. It seems interesting to investigate computational efficiency of the
approach in~\cite{Kornyak} in comparison with that based on \Gr bases.

\section{ACKNOWLEDGMENTS}

The research presented in this paper was partially supported by
the grant 04-01-00784 from the Russian Foundation for Basic
Research.



\begin{thebibliography}{1}

\bibitem{Dawson}
C.M. Dawson et al.
  {\em Quantum computing and polynomial equations over the finite field $Z_2$}.
   arXiv:quant-ph/0408129.

\bibitem{GS05}
V.P.Gerdt and V.M.Severyanov. {\em A Software Package to Construct Polynomial Sets over $Z_2$ for
Determining the Output of Quantum Computations.} arXiv:quant-ph/0509064.

\bibitem{Buch65} B.Buchberger. {\em An Algorithm for Finding a Basis for the
Residue Class Ring
of a Zero-Dimensional Polynomial Ideal}. PhD Thesis,
University of Innsbruck, 1965 (in German).

\bibitem{BW98}
B.Buchberger and F.Winkler (eds.) {\em Gr\"{o}bner Bases and Applications}. Cambridge University
Press, 1998.

\bibitem{Gerdt} V.P.Gerdt.
{\em Involutive Algorithms for Computing \Gr Bases}.
In: ``Computational commutative and non-commutative algebraic
geometry'', IOS Press, Amsterdam, 2005, pp. 199--225.
arXiv:math.AC/0501111.

\bibitem{Csharp} {\em Microsoft Visual C\# .net Standard}, Version 2003.

\bibitem{Shi} Y.Shi. {\em Both Toffoli and Control-Not need little help to do universal quantum computation},
Quantum Information and Computation, 3(1):84-92, 2003. arXiv:quant-ph/0205115.

\bibitem{Aharonov} D.Aharonov.
   {\em A Simple Proof that Toffoli and Hadamard Gates are Quantum Universal}.
     arXiv:quant-ph/0301040.

\bibitem{Nielsen} M.A.Nielsen and I.L.Chuang. {\em Quantum Computation and Quantum Information}. Cambridge
University Press, Cambridge, 2000.

\bibitem{Wang} D.Wang. {\em Elimination Methods}. Springer-Verlag, Berlin, 1999.

\bibitem{F4} J.-C. Faug\`{e}re. {\em A new efficient algorithm for computing Gro"bner bases (F4)},
Journal of Pure and Applied Algebra, 139(1-3): 61-88, 1999.

\bibitem{F5} J.-C.Faug\`{e}re. {\em A new efficient algorithm for computing \Gr bases without
reduction to zero ($F_5$)}. In: ``Proceedings of Issac 2002'', ACM Press, New York, 2002, pp.
75--83.

\bibitem{Gathen} J. von zur Garthen and J.Gerhard. {\em Modern Computer Algebra}. 2nd edition,
Cambridge University Press, 2003.

\bibitem{Inria} {\em Project-Team Spaces. Solving Problems through Algebraic
Computation and Efficient Software}. Activity Report, INRIA, 2003, p.13.
http://www.inria.fr/rapportsactivite/RA2003/

\bibitem{Kornyak} V.V.Kornyak. {\em On Compatibility of Discrete Relations}.
Lecture Notes in Computer Science. Springer-Verlag, Berlin, 2005, pp.272-284.
arXiv:org/math-ph/0504048.

\bibitem{ginv} http://invo.jinr.ru

\end{thebibliography}
\end{document}